\documentclass[aps,prl,twocolumn,showpacs,floatfix,nofootinbib,preprintnumbers,superscriptaddress,amsmath,amssymb]{revtex4-1}

\usepackage{epsfig}
\usepackage{color}
\usepackage{natbib}
\usepackage{graphicx}
\usepackage{xifthen}
\usepackage{bm}

\renewcommand{\vec}[1]{\mbox{\boldmath $#1$}}

\newcommand{\Sn}{$^{132}$Sn}
\newcommand{\Og}{$^{302}$Og}
\newcommand{\SHE}{$^{472}$164}
\begin{document}

\title{Electron and Nucleon Localization Functions of Oganesson: Approaching the Thomas-Fermi  Limit}

\author{Paul Jerabek}
\affiliation{%
Centre for Theoretical Chemistry and Physics, The New Zealand Institute for Advanced Study, Massey University Auckland, 0632 Auckland, New Zealand
}

\author{Bastian Schuetrumpf}
\affiliation{%
 NSCL/FRIB Laboratory, Michigan State University, East Lansing, Michigan 48824, USA
}

\author{Peter Schwerdtfeger}
\affiliation{%
Centre for Theoretical Chemistry and Physics, The New Zealand Institute for Advanced Study, Massey University Auckland, 0632 Auckland, New Zealand
}

\author{Witold Nazarewicz}
\affiliation{%
Department of Physics and Astronomy and FRIB Laboratory, Michigan State University, East Lansing, Michigan 48824, USA
}

\begin{abstract}
Fermion localization functions are used to discuss electronic and nucleonic shell structure effects in the superheavy element oganesson, the heaviest element discovered to date. Spin-orbit splitting in the $7p$ electronic shell becomes so large ($\sim$ 10 eV) that Og is expected to show  uniform-gas-like behavior in the valence region with a rather large dipole polarizability compared to the lighter rare gas elements. The nucleon localization in Og is also predicted to undergo a transition to the  Thomas-Fermi gas behavior in the valence region. This effect, particularly strong for neutrons, is due to the high density of single-particle orbitals.
\end{abstract}

\maketitle

\textit{Introduction --} Oganesson $(Z=118$) is the recent addition to the Periodic
Table of the Elements and the Chart of Nuclides \cite{Karol2016}. The isotope $^{294}_{118}$Og  was produced in a heavy ion fusion reaction with a $^{48}_{20}$Ca beam and a $^{249}_{98}$Cf target \cite{Oga1,Oga2006}. The heaviest element studied chemically to date is Fl ($Z=114$). Its relatively long half-life, 1-2\,s, enables chemical studies with $\sim$5 atoms/day, which marks
the limit of chemistry today \cite{Duellmann2017,Turler2015}.
The estimated $\alpha$-decay half-life of $^{294}_{118}$Og, 0.89$^{+1.07}_{-0.31}$ ms, is
too short for  chemical  ``one-atom-at-a-time" studies; hence,  its chemical properties must be  inferred from advanced atomic calculations based on relativistic quantum theory \cite{pitzer-1975,Pitzer-1975a,Pyyk88,Nash-1999,Goidenko2003,Nash2005,Pershina2008a,Pyyk11,Kullie-2012,Turler-Valeria-2013,Eliav-2015,Schwerdt2015,Pershina15,Shee-2015}.
According to these, Og has a closed-shell [Rn]$5f^{14}6d^{10}7s^{2}7p^{6}$ configuration~\cite{Desclaux-1973,Indel07,Pyyk11}, with
a very large spin-orbit splitting of the $7p$ shell (9.920 eV at the Dirac-Breit-Hartree-Fock and 10.125 eV at the Fock-Space Coupled-Cluster level, see below).
While, according to its electronic configuration (Og 
completes the 7th row of the Periodic Table), it does not behave like a typical rare gas element.
For example, the relativistic $7p_{3/2}$ expansion and the relativistic $8s$ contraction make Og the first rare gas element with a positive electron affinity of 0.064 eV \cite{Eliav-2015,Eliav-1996,Goidenko2003}. This result includes a substantial quantum electrodynamic correction of 0.006 eV \cite{Eliav-2015}.

Nuclear structure calculations predict $^{294}$Og to be a deformed nucleus \cite{Cwiok-2005,Erler12,Heenen2015,Schuetrumpf2017a}, 
eight neutrons away from  the next neutron shell closure at $^{302}$Og ($N=184$)~\cite{cwiok96,Bender1999,Bender2001,Berger03,Afa05,Agb15}. A new factor 
impacting properties of superheavy nuclei 
is the strong electrostatic repulsion: the Coulomb  force in superheavy nuclei cannot be treated as a small perturbation atop the dominating nuclear interaction; the resulting polarization effects due to Coulomb frustration are expected to  influence significantly proton and neutron distributions and shell structure \cite{Moeller1992,Bender1999,Decharge1999,Berger2001,Decharge2003,Afa05,Pei2005,Schuetrumpf2017a}.
In particular, the isotope $^{294}_{118}$Og is believed to be
a semi-bubble system  with a sizable central depression of the proton density \cite{Schuetrumpf2017a}.

The objective of this paper is to study the electronic and nucleonic shell structure of superheavy elements. The electronic shell structure is expected to be impacted by the transition from the $LS$-coupling of the Schr\"odinger equation at lower atomic numbers  to the $jj$-coupling  of the Dirac equation at large $Z$-values. 
In the nuclear case, the shell structure is expected to be washed out due to the large density of single-nucleonic states. While the kinematics of protons and neutrons in a nucleus   is non-relativistic, the large spin-orbit coupling (that is about an order of magnitude greater than in atomic case due to  large spin-dependent components of the nucleon-nucleon interaction \cite{Ring00}) results in a $jj$-coupling. Therefore, both for electronic and nucleonic systems, the pattern of single-particle levels  of superheavy species is expected to be strongly impacted by both radial and total angular momentum characteristics \cite{Boh75,Bra97}.
To describe these changes quantitatively, we utilize  the fermion localization measure \cite{Becke1990},  which is an excellent indicator of  shell structure. In particular, we investigate the transition from 
the regime of strong localization, governed by  shell effects, to a more delocalized regime typical of a uniform-density  Thomas-Fermi gas. As we shall demonstrate, superheavy species constitute an excellent territory where to look for such a transition.

\textit{Fermion localization function --}
The spatial localization measure was originally proposed in atomic and molecular physics to characterize shell structure and chemical bonding  in electronic systems \cite{Becke1990,Kohout96,savin1997,scemama2004,Kohout04,burnus2005,Poater}. It has been subsequently introduced to nuclear physics to visualize cluster structures in light nuclei \cite{Reinhard2011}. The novel nuclear applications include description of  nuclear fission \cite{Zhang2016} and heavy-ion fusion \cite{Schuetrumpf2017b}, and  nucleonic matter in the inner crust of neutron stars \cite{Schuetrumpf2017}. In electronic systems, spatial localization function is referred to as the electron localization function (ELF); in nuclear systems as the nucleon localization function (NLF). It is based on the inverse of the conditional probability of finding a fermion of type $q$ ($=e$, $n$, or $p$) in the vicinity of another fermion of the same type and same spin/signature quantum number $\sigma$ ($=\uparrow$ or $\downarrow$), knowing that the latter particle is located at position $\vec{r}$. 
While this  probability is generally  given by the non-local one-body density matrix \cite{Becke1990}, it is useful to introduce a local quantity that provides information about the short-range behavior. To this end, Becke and  Edgecombe \cite{Becke1990} introduced 
the local  measure of fermion localization, which -- in the non-relativistic case -- can be written as:
\begin{equation} 
\mathcal{C}_{q\sigma}(\vec{r})=\left[1+\left(\frac{\tau_{q\sigma}\rho_{q\sigma}-\frac{1}{4}|\vec{\nabla}\rho_{q\sigma}|^2-\vec{j}^2_{q\sigma}}{\rho_{q\sigma}\tau^\mathrm{TF}_{q\sigma}}\right)^2\right]^{-1},
\label{eq:localization}
\end{equation}
where $\rho_{q\sigma}$, $\tau_{q\sigma}$, $\vec{j}_{q\sigma}$, and $\vec{\nabla}\rho_{q\sigma}$ are the particle density, kinetic energy density, current density, and density gradient, respectively. $\tau^\mathrm{TF}_{q\sigma}$ denotes the Thomas-Fermi kinetic energy. In this work, time reversal symmetry is conserved; hence, $\vec{j}_{q\sigma}$ vanishes.

The localization function takes generally values between 0 and 1. A value close to 1 indicates that the probability of finding two particles (of the same type) close to each other is very low. Thus a high value of $\mathcal{C}$ marks the spatial regions corresponding to shell separations. 
Since the localization function (\ref{eq:localization}) is normalized to the Thomas-Fermi kinetic energy, $\mathcal{C}=1/2$ corresponds to the limit of the uniform-density Fermi gas, in which the individual orbits are spatially delocalized.

\begin{figure}[tb]
 \includegraphics[width=0.9\linewidth]{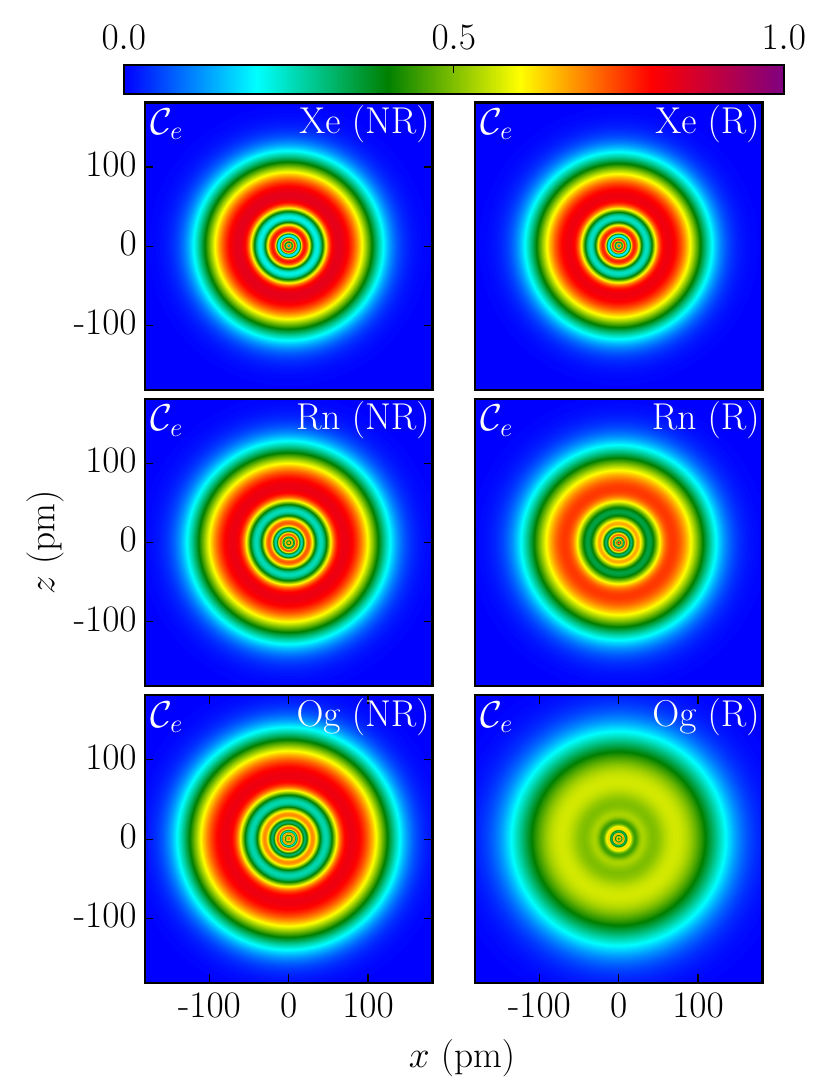}
\caption{ELFs from nonrelativistic (NR, left) and Dirac-Hartree-Fock calculations (R, right) for the heavy rare gas atoms Xe (top), Rn (middle), and Og (bottom).}
\label{fig:nr-rel-elf}
\end{figure}
\textit{Electron localization --} For the electronic structure calculations we used the {\sc elf} module as implemented in the relativistic ab-initio quantum chemistry program {\sc dirac15} \cite{Dirac15}.  Hartree-Fock one-particle densities were generated in non-relativistic, scalar-relativistic (module {\sc x2c}-spinfree) \cite{Dyall1994,Ilias2007}, and (4-component) Dirac-Coulomb calculations in conjunction with an uncontracted  relativistic quadruple-zeta basis set {\sc dyall.acv4z} \cite{Dyall2006}. The Dirac-Fock computations include the small-component integrals as well as the two-electron Gaunt term. We utilized the finite-field method to compute the static electric dipole polarizability of Og (with external electric field strengths of 0.0, 0.0005 and 0.001~a.u.) at CCSD(T) Coupled-Cluster level,\cite{Pershina2008a} which included excitations from singles, doubles, and perturbative triples. In the correlation treatment, we included  50 electrons and virtual orbitals up to 25~a.u. Here we used the molecular mean-field {\sc x2c} Hamiltonian \cite{Sikkema2009} with the Gaunt term included. Fock-Space Coupled-Cluster calculations \cite{Eliav-2015} were carried out to obtain the ionization potentials from the filled 7$p_{3/2}$ and 7$p_{1/2}$ shells of Og. Note that only large-component densities are considered for the non-relativistic and scalar-relativistic ELF, whereas in the 4-component case the small-component densities are added to the large-components to yield the total one-particle density. Relativistic effects make a huge imprint on many properties of Og. For instance, the electron binding energy of Og is predicted to rise by as much as 227\,keV by considering relativistic effects (for comparison, a similar number for Pb is a mere 40\,keV).

\begin{figure}[tb]
\includegraphics[width=0.7\linewidth]{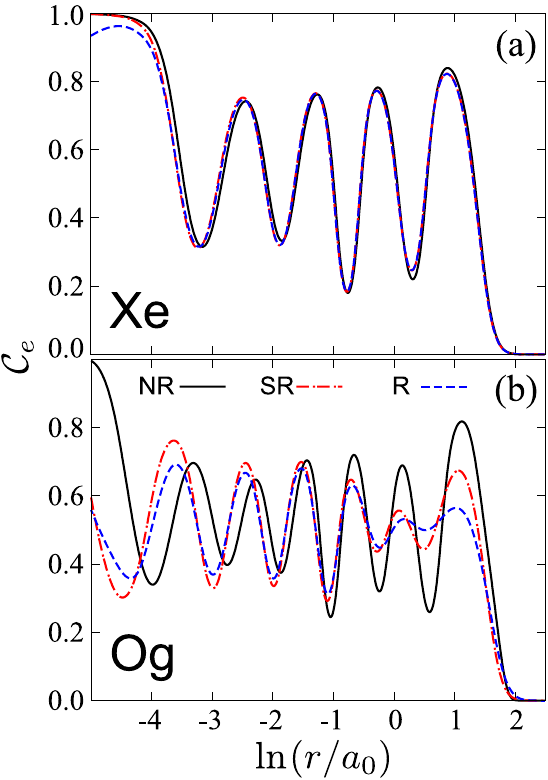} 
\caption{ELFs for Xe (a) and Og (b) from non-relativistic (NR), scalar-relativistic (SR), and Dirac-Hartree-Fock (R) calculations as a function of the distance from the nucleus as in Ref.~\cite{Becke1990}. The relativistic contraction of inner shells  and smearing out of the shell structure in the valence and sub-valence shells  of Og are clearly seen.}
   \label{fig:xe-og-1d-elf}
\end{figure}

Figures~\ref{fig:nr-rel-elf} and  \ref{fig:xe-og-1d-elf} show the ELFs predicted in our calculations.
As seen in  Fig.~\ref{fig:nr-rel-elf}, electron localizations 
for Xe or Rn hardly change from the nonrelativistic to the 4-component relativistic framework. However, for Og we see significant electron delocalization with ELF values that are much smaller compared to the non-relativistic case, making the atomic shell structure barely recognizable.
The pattern of concentric rings is a fingerprint of the underlying shell structure. The  sizes
of rings in ELF 
reflect the radii of electron orbits in different shells; hence, they roughly scale with  $n^2$, where $n$ is the principal quantum number \cite{Becke1990,Kohout96}.
Figure \ref{fig:xe-og-1d-elf}(b) clearly shows that the delocalization is mainly due to spin-orbit coupling and not due to scalar relativistic effects. This results in  an evenly distributed ELF with values around 0.5 in the outer shells. The valence and sub-valence shells of Og are, therefore, smeared out like in a homogenous electron gas. Rn behaves similarly to Xe, although some delocalization through relativistic effects is already apparent. 

\begin{figure}[htb]
  \includegraphics[width=\linewidth]{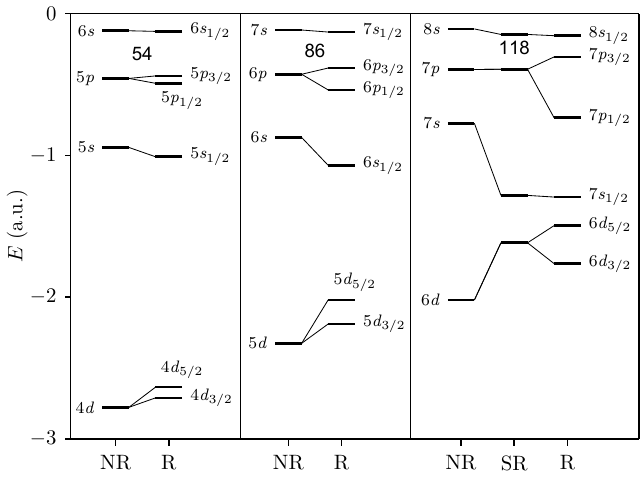}
   \caption{Orbital energy levels of Xe (left), Rn (middle), and Og (right) for the $^1S_0$ ground state as obtained from non-relativistic (NR) and scalar-relativistic (SR) Hartree-Fock and Dirac-Hartree-Fock (R) calculations. $6s$ (Xe), $7s$ (Rn), and $8s$ (Og) orbital energies taken from the first excited $^3P_2$ state.}
     \label{fig:orbitalenergies}
 \end{figure}

A more detailed analysis shows that smearing out of the electron density in the valence region originates from the  strong spin-orbit splitting of the 7$p$ shells; while the radii  for the valence 5$p$ orbitals in Xe are very similar (2.239 and 2.141~a.u. for 5$p_{3/2}$ and 5$p_{1/2}$, respectively, as obtained with the numerical program {\sc grasp92} \cite{GRASP}) the 7$p_{3/2}$ shell in Og is 0.75~a.u. further out compared to the 7$p_{1/2}$ shell (2.796 and 2.039~a.u., respectively). Large spin-orbit splittings are also calculated for the lower lying $\ell>0$ (core) shells. Further, the density of the single-particle (s.p.) states increases from Xe to Og as expected for higher principal quantum numbers, see Figure~\ref{fig:orbitalenergies}. 
As a result of these effects, the electron density is more homogeneously distributed over the entire atomic range, i.e., less localized, resulting in ELF values oscillating around the Thomas-Fermi limit. Our Fock-Space Coupled-Cluster calculation gave ionization potentials  of 7$p_{3/2}$ and 7$p_{1/2}$ of 8.842 and 18.967 eV, respectively, thus spin-orbit splitting for the valence $7p$ orbital of Og is extremely large (10.125~eV). Figure~\ref{fig:orbitalenergies} illustrates this in relation to the orbital energy levels of the lighter homologues.

According to the Thomas-Fermi model, the static dipole polarizability  $\alpha\propto r_a^3$, with $r_a^3$ being the atomic radius \cite{Shevelko-1979}. 
Our state-of-the-art calculations  show that
the electron-gas-like  outer shell of Og,
resulting in  $\alpha = 57.98$~a.u., is  much easier to polarize as compared to xenon ($\alpha = 27.815$~a.u.~\cite{Hohm1990}) or radon ($\alpha = 33.18$~a.u.~\cite{Nakajima2001}). For comparison, the nonrelativistic and scalar relativistic values for Og are $\alpha = 45.30$~a.u. and $\alpha = 43.78$~a.u., respectively. Thus, for Og one expects an increase in van-der-Waals interactions compared to the lighter rare gases, and subsequently a significant change in chemical and physical properties of this element, see also Refs.~\cite{Pershina2008a,Goidenko2003,Nash2005,Hangele-2013,Turler-Valeria-2013,Shee-2015} for more discussion on this point.

\textit{Nucleon localization --} 
For the nuclear calculations, we employ nuclear density functional theory \cite{bender2003self} with carefully optimized global Skyrme energy density functionals   UNEDF1 \cite{Kortelainen2012} and  SV-min \cite{Kluepfel2009}. 
Pairing is of minor importance in the closed-shell nuclei considered. It is treated as in Ref.~\cite{Schuetrumpf2017a}. Namely, we consider the density-dependent contact force at the level of BCS theory. The pairing space is limited by a soft cutoff with the cutoff parameter chosen such that it covers about 1.6 extra oscillator shells above the Fermi energy.  We use the DFT solver of Ref.~\cite{Rei91aR} constrained to spherical geometry as all nuclei considered are expected to be spherical in their ground states.

 \begin{figure}[tb]
  \includegraphics[width=0.9\linewidth]{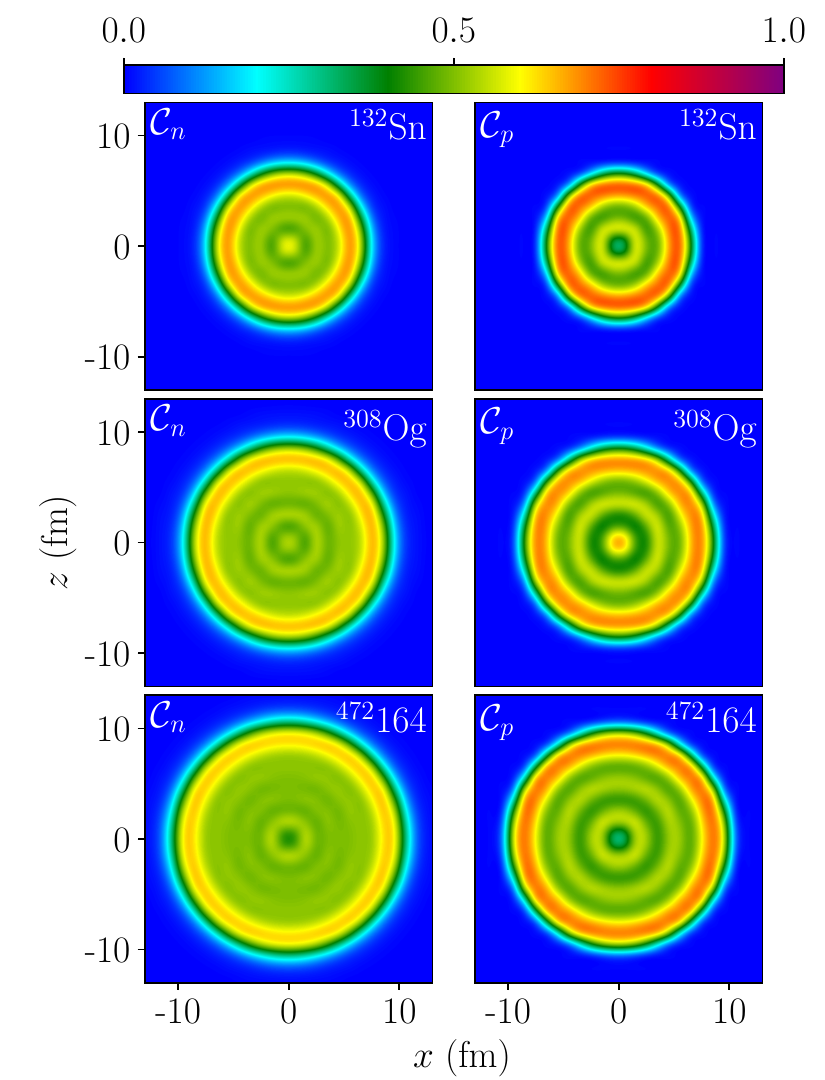}
   \caption{NLFs of \Sn{}, \Og{}, and \SHE{} calculated with the energy density functional UNEDF1.}
     \label{fig:nuclear_loc}
 \end{figure}
Figure~\ref{fig:nuclear_loc} shows the NLFs for the doubly magic medium-mass nucleus \Sn{} and spherical superheavy systems \Og{}  and \SHE{}. We consider the latter ``theoretical" nucleus to further illustrate the behavior of NLFs at still larger numbers of nucleons.
In contrast to the ELFs, 
the number of closed shells cannot be determined from the number of radial maxima. This is due to the different radial behavior of single-nucleon orbits. While the radii of electron orbits in atoms belonging to different shells are spatially well separated,  radii of nucleonic 
orbits   scale roughly  as $\sim \sqrt{2n_r+\ell}$, i.e., they  very gradually increase with the shell number. This results in a large spatial overlap between single-nucleon wave functions and reduced localizations as compared to the electronic case.  A characteristic  feature of  NLFs is the local enhancement  at the surface  \cite{Zhang2016} due to the fact that  few  valence  nucleons contribute to the total density at distances greater than the nuclear radius.

Inspecting the NLFs of protons to neutrons, one notes that the patterns of concentric rings is more distinct in the proton system, as the number of occupied proton shells is less than that for the neutrons, within the same volume (as the rms proton and neutron radii are very similar \cite{Erler12}).
This effect becomes  fairly pronounced for superheavy nuclei where the neutron excess is large.
While the NLF for the medium-mass nucleus \Sn{} exhibits a clear shell structure with distinct oscillations around $\mathcal{C}=0.5$ \cite{Zhang2016}, the maxima and minima become fainter for  heavier systems. This is particularly striking for the neutrons. While the neutron NLF for \Og{} still exhibits  a faint structure in the interior, the ring pattern  almost vanishes for \SHE{}.  Overall, as mass increases,  the neutron localization  approaches the Thomas-Fermi  limit   $\mathcal{C}=0.5$ in the valence region ($r>3$\,fm) below the surface peak.
The NLF pattern seen in Fig.~\ref{fig:nuclear_loc} reflects the underlying nucleonic shell structure. As discussed in, e.g., Refs.~\cite{cwiok96,Bender1999,Bender2001,Berger03,Afa05,Agb15} the general pattern of s.p. energies undergoes significant changes in superheavy nuclei. First,  the  s.p. level density is large; in fact it  grows faster than $A^{1/3}$ \cite{Agb15}. Consequently -- similar to what has been  discussed earlier in the context of atomic calculations of the electron shell structure of superheavy elements -- small changes in the theoretical description   can impact shell structure substantially.  Second, the shell structure of superheavy nuclei is influenced by
the self-consistent interplay between the short-range attractive nuclear force and the long-range electrostatic repulsion. Thanks to the resulting  Coulomb frustration, significant rearrangements of nucleonic densities, such as the appearance of central depression, are predicted \cite{Moeller1992,Bender1999,Decharge1999,Berger2001,Decharge2003,Afa05,Pei2005,Schuetrumpf2017a}.
The presence  of central depression strongly affects high-$j$ orbits due to their large s.p. radii
\cite{cwiok96,Decharge1999,Decharge2003,Afa05}.

\textit{Conclusions --} To study electronic and nucleonic shell structure in superheavy elements, we employed  the  local spatial  measure of fermion localization. The atomic calculations were carried out for heavy rare gas atoms Xe, Rn, and the superheavy element  Og recently added to the Periodic Table. The nuclear calculations were performed for the known doubly-magic system \Sn{} and for superheavy nuclei \Og{}, and \SHE{}. This study constitutes the first application of fermion localization to superheavy atoms and nuclei.

Relativistic effects significantly impact the electronic structure of superheavy atoms.
For the  element Og, the electron shells  with $\ell>0$ show very large spin-orbit splittings smearing out of the one-particle density, thus becoming more uniformly distributed over the entire atom approaching  the electron-gas regime in the valence region. A direct consequence of this transition is its predicted large static dipole polarizability resulting in 
an increase in van-der-Waals interactions compared to the lighter rare gases and  a significant change in its chemical and physical properties.

A gradual transition towards the uniform-gas regime is predicted for  nucleonic localizations in superheavy nuclei. In general, neutrons are more delocalized than protons as for the superheavy nuclei $N$ is much greater than $Z$, i.e., more neutrons are confined to the same volume than protons. While the semiclassical Thomas-Fermi limit in nuclei is strictly approached only for systems with extremely large particle numbers $A > 5000$  \cite{Bra97,Bohigas76,Reinhard-2006}, we can see that in the discussed superheavy nuclei the Fermi-gas limit of neutron NLFs is reached 
in the valence region ($r>3$\,fm) below the surface peak.
 
In summary, through electron and nucleon localization functions we show that Og is a rather unusual addition to the Periodic Table and to the Chart of Nuclides. High density of electronic and nucleonic s.p. states, relativistic effects resulting in the strong spin-orbit splitting of electronic levels, and nucleonic polarization effects, make the superheavy atoms, such as Og, quantitatively different from the lighter congeners.

\begin{acknowledgements}
We acknowledge financial support by the Alexander-von-Humboldt Foundation (Bonn, Germany) and the
Marsden Fund of the Royal Society of New Zealand.  This work was also supported by the U.S. Department of Energy,  under Award Numbers DOE-DE-NA0002847 (NNSA, the Stewardship Science Academic Alliances program) and  DE-SC0013365 and DE-SC0008511 (Office of Science).
\end{acknowledgements}

\bibliographystyle{apsrev4-1}
\bibliography{nucllib}

\end{document}